\documentclass [aps,preprint,prd,showpacs,nofootinbib,superscriptaddress]{revtex4}
\usepackage{verbatim}
\usepackage{slashed}
\usepackage{epsfig}
\usepackage{feynmf}
\usepackage{graphicx}
\usepackage{amsmath}
\usepackage{mathrsfs}
\usepackage{bm}

\begin{document}


\title{Predictions of Light Hadronic Decays of Heavy Quarkonium ${}^1D_2$ States in NRQCD}


\author{Ying \surname{Fan}}
\email{ying.physics.fan@gmail.com}
\author{Zhi-Guo He}
\email{hzgzlh@gmail.com}
\author{Yan-Qing Ma}
\email{yqma.cn@gmail.com}
\author{Kuang-Ta Chao}
\email{ktchao@th.phy.pku.edu.cn} \affiliation{Department of Physics
and State Key Laboratory of Nuclear Physics and Technology, Peking
University, Beijing 100871, China}




\date{\today}

\begin{abstract}
The inclusive light hadronic decays of ${}^1D_2$ heavy quarkonia are
studied within the framework of NRQCD at the leading order in $v$
and up to the order of $\alpha_s^3$. With one-loop QCD corrections,
the infrared divergences and Coulomb singularities in the decay
amplitudes are proved to be absorbed by the renormalization of the
matrix elements of corresponding NRQCD operators, and the infrared
finite short-distance coefficients are obtained through the matching
calculations. By taking the factorization scale to be $2m_Q$, the
light hadronic decay widths are estimated to be about 274, 4.7, and
8.8 KeV for the $\eta_{c2},~ \eta_{b2}$, and $\eta_{b2}'$
respectively. Based on the above estimates, and using the E1
transition width and dipion transition width for the $\eta_{c2}$
estimated elsewhere, we get the total width of $\eta_{c2}$ to be
about 660-810~KeV, and the branching ratio of the  E1 transition
$\eta_{c2}\to\gamma\,h_c$ to be about $(44\,\mbox{-}\,54)\%$, which
will be useful in searching for this missing charmonium state
through, e.g., the process $\eta_{c2}\to\gamma\,h_c$ followed by
$h_c\to\gamma\eta_c$.


\end{abstract}


\maketitle




\section{Introduction}
The studies of production and decay mechanisms for heavy quarkonia
provide important information on both perturbative and
nonperturbative QCD. Based on the nonrelativistic (NR) nature of
heavy quarkonium systems, an effective field theory, the
nonrelativistic QCD (NRQCD) factorization formulism was proposed by
Bodwin, Braaten and Lepage in 1990s\cite{BBL}. Within this
framework, the inclusive decay and production of heavy quarkonium
can be factorized into two parts, the short distance coefficients
and the long distance matrix elements.
Differing from the color-singlet model (CSM)~\cite{CSM}, in the
NRQCD factorization formalism, the heavy quark and antiquark pair
annihilated or produced at short distances can be in both the
color-singlet and the color-octet states with the same or different
angular momentum quantum numbers~\cite{BBL}, and the latter is known
as the color-octet mechanism (COM).
This mechanism has been used to remove the infrared divergences in
inclusive P-wave charmonium production~\cite{Pproduction} and
decay~\cite{BBL,Huang1,Huang2,Huang3} to give the infrared safe and
model independent predictions.

Recently, the inclusive light hadronic decays of ${}^3D_J$
charmonium states were also studied within the framework of NRQCD
factorization up to order $\alpha_s^3$\cite{Heshort,He}. The
infrared divergence found in the CSM
calculation~\cite{Dwave-decay:CSM} is removed by absorbing it into
the matrix elements of the color-octet ${}^3P_J$ operators.
Furthermore, the new contributions at order $\alpha_s^2$ from the
color-octet ${}^3P_J$ and ${}^3S_1$ matrix elements enhance the
decay widths of ${}^3D_J$ states, and the numerical results are
larger than those estimated in the CSM by several times in
magnitude~\cite{He}. One can expect that a similar case will emerge
in the inclusive light hadronic decay of ${}^1D_2$ charmonium,
namely, the $\eta_{c2}$ state. The difference between the
$\eta_{c2}$ and ${}^3D_J$ states is that there are no infrared
divergences in the inclusive decay width of $\eta_{c2}$ in the CSM
up to order $\alpha_s^2$~\cite{Novikov78}, and the numerical result
is about 110 KeV~\cite{Eichten}. However, the infrared divergence
will emerge again in the decay width of $\eta_{c2}$ in the CSM at
order $\alpha_s^3$, which needs to be removed by invoking the
color-octet mechanism, i.e. by absorbing it into the corresponding
color-octet matrix elements.

On the other hand, the estimation of the inclusive light hadronic
decay width of $\eta_{c2}$ is also important phenomenologically for
probing this missing charmonium state. Quark model predicts its mass
within the range 3.80-3.84 GeV~\cite{Eichten04:spectrum,Barnes},
which lies between the $D\bar{D}$ and the $D^*\bar{D}$ thresholds.
However, its odd parity ($J^{PC}=2^{-+}$) forbids the decay to
$D\bar{D}$. As a result, it should be a narrow state, and its main
decay modes are the electric as well as hadronic transitions to
lower-lying charmonium states and the inclusive light hadronic
decay. Therefore, the study for the inclusive light hadronic decay
of $\eta_{c2}$ in NRQCD factorization will provide important
information on searching for this state in high-energy $p\bar p$
collision~\cite{Cho}, in $B$ decays~\cite{Ko}, in higher charmonium
transitions, and in the low-energy $p\bar p$ reaction in PANDA at
FAIR~\cite{PANDA} and in $e^+e^-$ process in BESIII at
BEPC~\cite{BES3}.


In this paper, we study the one-loop QCD corrections to light
hadronic decay of ${}^1D_2$ within the framework of NRQCD
factorization. The paper is organized as follows: after an
introduction of the NRQCD factorization formalism in Sec.~II, we
calculate the decay widths up to $\mathcal{O}(\alpha_s^3)$ in
perturbative QCD in Sec.~III, where both the real and virtual
corrections are considered. Then perturbative NRQCD is applied to
obtain the imaginary parts of the forward scattering amplitudes in
Sec.~IV. Combined with the QCD results, the infrared divergences are
either canceled or absorbed into the long distance NRQCD matrix
elements, and the finite short distance coefficients are obtained.
Together with the long distance matrix elements estimated by solving
the operator evolution equations, the decay width is determined. The
numerical results and phenomenological discussions are given in
Sec.~V. In the last, we will give a brief summary for our results in
Sec.~VI.

\section{General Formulas}

There are four important scales in the heavy quarkonium system: the
heavy quark mass $m_Q$,  the typical momentum of the heavy quark or
the inverse of the size of the bound state $m_Q\,v$~\footnote{Here,
$v$ denotes the relative velocity of the heavy quark pair in the
meson frame. The average value of $v^2$ is about 0.3 for charmonium
and about 0.1 for bottomonium~\cite{BBL}.}, the binding energy
$m_Q\,v^2$ and the QCD scale $\Lambda_{QCD}$, while the dynamical
property of the bound state is mainly determined by the latter three
scales. Thus one can choose a cutoff $\mu_{\Lambda}$ with condition
$m_Q>\mu_{\Lambda}\gg m_Q\,v\ (m_Q\,v^2,\ \Lambda_{QCD})$ to
integrate out the hard scale $m_Q$. Expanding the nonlocal effective
action in power of $v$ and writing the result in the two-component
Pauli spinor space, one then can get the effective Lagrangian for
NRQCD~\cite{BBL}:
\begin{equation}
\mathcal{L}_{NRQCD}=\mathcal{L}_{light}+\mathcal{L}_{heavy}+\delta\mathcal{L},
\label{L:NRQCD}\end{equation} where the Lagrangian
$\mathcal{L}_{light}$ describes gluon and light quarks. At leading
order in $v$, the heavy quark and antiquark are described by
$\mathcal{L}_{heavy}$:
\begin{equation}\label{Lag}
\mathcal{L}_{heavy}=\psi^{\dagger}(iD_{t}+\frac{\mathbf{D}^{2}}{2m_{Q}})\psi+
\chi^{\dagger}(iD_{t}-\frac{\mathbf{D}^{2}}{2m_{Q}})\chi,
\end{equation} where $\psi$ denotes the Pauli spinor
field that annihilates a heavy quark, $\chi$ denotes the Pauli
spinor field that creates a heavy antiquark, and $D_t$ and
$\mathbf{D}$ are the time and space components of the
gauge-covariant derivative $D^{\mu}$, respectively. The relativistic
corrections to $\mathcal{L}_{heavy}$ are included in the term
$\delta\mathcal{L}$. The most important correction terms for heavy
quarkonium energy splitting are the bilinear ones:
\begin{eqnarray}
\delta{\mathcal{L}}_{bilinear}&=&\frac{c_{1}}{8m_{Q}^3}[\psi^{\dagger}(\mathbf{D}^2)^{2}\psi-
\chi^{\dagger}(\mathbf{D}^2)^{2}\chi]\nonumber\\
&+&\frac{c_{2}}{8m_{Q}^2} [\psi^{\dagger}(\mathbf{D}\cdot
g\mathbf{E}-g\mathbf{E}\cdot
\mathbf{D})\psi+\chi^{\dagger}(\mathbf{D}\cdot
g\mathbf{E}-g\mathbf{E}\cdot \mathbf{D})\chi]\nonumber \\
&+&\frac{c_{3}}{8m_{Q}^2}[\psi^{\dagger}(i\mathbf{D}\times
g\mathbf{E}-g\mathbf{E}\times
i\mathbf{D})\cdot\boldsymbol{\sigma}\psi+\chi^{\dagger}(i\mathbf{D}\times
g\mathbf{E}-g\mathbf{E}\times
i\mathbf{D})\cdot\boldsymbol{\sigma}\chi]\nonumber\\
&+&\frac{c_{4}}{2m_{Q}}[\psi^{\dagger}(g\mathbf{B}\cdot\boldsymbol{\sigma})\psi-
\chi^{\dagger}(g\mathbf{B}\cdot\boldsymbol{\sigma})\chi],
\label{L:bilinear}\end{eqnarray} where $E^i=G^{0i}$ and
$B^i=\frac{1}{2}\epsilon^{ijk}G^{jk}$ are the electric and magnetic
components of the gluon field-strength tensor $G^{\mu\nu}$,
respectively.

In the Lagrangian $\mathcal{L}_{NRQCD}$ in (\ref{L:NRQCD}), there
are still three low-energy sales: the soft scale $m_Q\,v$, the
ultrasoft scale $m_Q\,v^2$ and the QCD scale $\Lambda_{QCD}$. The
existence of multi-scales makes the power counting rules of NRQCD
(the velocity scaling rules~\cite{BBL}) can not be homogeneous
generally. More seriously, if one wants to do the NRQCD loop
calculations in dimensional regularization scheme with
$\mathcal{L}_{NRQCD}$ defined in (\ref{L:NRQCD}), one will find that
the hard scale can not decouple from the loop integrals and the
power counting rules are violated inevitably~\cite{Manohar97:NRQCD}.
These problems can be solved simultaneously by the method of
regions~\cite{Beneke}, which will be explained and applied in our
calculations in Sec.~IV.

To reproduce the annihilation contribution to a low-energy
$Q\bar{Q}\rightarrow Q\bar{Q}$ scattering amplitude in NRQCD, local
four-fermion operators in $\delta\mathcal{L}$ are needed, which have
the general form~\cite{BBL}
\begin{equation}
\delta\mathcal{L}_{4-fermion}=\sum_{n}\frac{f_{n}(\mu_\Lambda)}{m_{Q}^{d_{n}-4}}\mathcal{O}_{n}(\mu_\Lambda).
\end{equation}
where $\mathcal{O}_{n}$ denotes regularized local four-fermion
operators, such as $\psi^{\dagger}\chi\chi^{\dagger}\psi$, and $d_n$
is the naive scaling dimension of the operator. The dependence on
cutoff $\mu_\Lambda$ of the operator $\mathcal{O}_{n}$ is canceled
by that of scaleless coefficient $f_{n}(\mu_\Lambda)$, which can be
computed by matching the full QCD onto the NRQCD as perturbation
series in $\alpha_s$.

In NR theory, the width of heavy quarkonium $H$ is $-2$ times the
imaginary part of the energy of the state, thus one has~\cite{BBL}
\begin{equation}
\Gamma(H\rightarrow LH)=2\textrm{Im}\langle
H|\delta\mathcal{L}_{4-fermion}|H\rangle=\sum_{n}\frac{2\textrm{Im}f_{n}(\mu_\Lambda)}{m_{Q}^{d_{n}-4}}\langle
H|\mathcal{O}_{n}(\mu_\Lambda)|H\rangle,
\label{Gamma:HtoLH}\end{equation} where $LH$ represents all possible
light hadronic final states, and the operator $\mathcal{O}_{n}$ here
and afterward only denotes the one relevant to the strong
annihilation of $Q\bar{Q}$. The NR normalization has been applied
for the state $|H\rangle$ in (\ref{Gamma:HtoLH}).

In order to calculate the coefficients of four-fermion operators in
(\ref{Gamma:HtoLH}), the equivalence of full QCD and NRQCD at long
distance is exploited. Since in construction, the coefficient
$f_{n}$ is of short-distance nature and is independent on the long
distance asymptotic state, one can get it by replacing the state
$|H\rangle$ by the on-shell heavy quark pair state
$|Q\overline{Q}\rangle$ with small relative momentum and matching
the forward scattering amplitude of $Q\overline{Q}\to Q\overline{Q}$
in full QCD onto that of NRQCD perturbatively. The matching
condition is written as~\cite{BBL}
\begin{equation} \label{eqm}
\mathcal{A}(Q\overline{Q}\rightarrow
Q\overline{Q})\Big{|}_{\textrm{pert
QCD}}=\sum_{n}\frac{f_{n}(\mu_\Lambda)}{m_Q^{d_{n}-4}}\langle
Q\overline{Q}|\mathcal{O}_{n}
(\mu_\Lambda)|Q\overline{Q}\rangle\Big{|}_{\textrm{pert NRQCD}}\,.
\end{equation}
Since we only need the imaginary parts of the coefficients, optical
theorem can be used to simplify the matching calculations.

The physical ${}^1D_2$ state can be expanded in powers of $v$ in the
Fock space:
\begin{equation}
|^1D_{2}\rangle=\mathcal{O}(1)| Q\bar{Q}(^1D_{2}^{[1]})\rangle+
\mathcal{O}(v)| Q\bar{Q}(^1P_{1}^{[8]})\rangle+ \mathcal{O}(v^{2})|
Q\bar{Q}(^1S_{0}^{[1,8]})\rangle+\mathcal{O}(v^3),\label{FockState}
\end{equation}
where the superindices [1] and [8] denote the color-singlet and
color-octet, respectively. The contributions from the P-wave and
S-wave Fock states to the annihilation rate of ${}^1D_2$  are at the
same order of $v^2$ as that from the D-wave state, because their
relevant operators scale $v^{-2}$ and $v^{-4}$ relative to
$\mathcal{O}_{1}(^{1}D_{2})$, as can be seen later. Other Fock
states contribute at higher order of $v^2$. Therefore the light
hadronic decay width of ${}^1D_2$ at leading order in $v^2$ can be
described in NRQCD factorization framework as follows:
\begin{eqnarray}\label{eq1}
&\Gamma&(^{1}D_{2}\rightarrow LH)=\nonumber\\
&2&\textrm{Im}f(^{1}D_{2}^{[1]})\frac{\langle
^{1}D_{2}|\mathcal{O}_{1}(^{1}D_{2})|^{1}D_{2}\rangle}{m_{Q}^{6}}
+2\textrm{Im}f(^{1}P_{1}^{[8]})\frac{\langle
^{1}D_{2}|\mathcal{O}_{8}(^{1}P_{1})|^{1}D_{2}\rangle}{m_{Q}^{4}}+\nonumber\\
&2&\textrm{Im}f(^{1}S_{0}^{[8]})\frac{\langle ^{1}D_{2}|
\mathcal{O}_{8}(^{1}S_{0})|^{1}D_{2}\rangle}{m_{Q}^{2}}+
2\textrm{Im}f(^{1}S_{0}^{[1]})\frac{\langle ^{1}D_{2}|
\mathcal{O}_{1}(^{1}S_{0})|^{1}D_{2}\rangle}{m_{Q}^{2}},
\end{eqnarray}
where the four-fermion operators are~\cite{Maltoni}:
\begin{eqnarray}
\mathcal{O}_{1}(^1S_{0})&=&\frac{1}{2N_{c}}\psi^{\dagger}\chi\chi^{\dagger}\psi,\nonumber\\
\mathcal{O}_{8}(^1S_{0})&=&\psi^{\dagger}T^{a}\chi\chi^{\dagger}T^{a}\psi,\nonumber\\
\mathcal{O}_{8}(^1P_{1})&=&\psi^{\dagger}(-\frac{i}{2}\overleftrightarrow{\boldsymbol{D}})T^{a}\chi\cdot
\chi^{\dagger}(-\frac{i}{2}\overleftrightarrow{\boldsymbol{D}})T^{a}\psi,\nonumber\\
\mathcal{O}_{1}(^{1}D_{2})&=&\frac{1}{2N_{c}}\psi^{\dagger}S^{ij}\chi
\chi^{\dagger}S^{ij}\psi, \end{eqnarray} where
$\overleftrightarrow{\boldsymbol{D}}=\overrightarrow{\boldsymbol{D}}-\overleftarrow{\boldsymbol{D}}$
and $S^{ij}=(-\frac{i}{2})^{2}
(\overleftrightarrow{D}^{i}\overleftrightarrow{D}^{j}-\frac{1}{3}\overleftrightarrow{\boldsymbol
D}^{2}\delta^{ij})$. Since $\boldsymbol{D}^{2}/m_Q^2$ scales as
$v^2$, it can be ensured that the four terms in (\ref{eq1}) are at
the same order of $v$.

The coefficients in (\ref{eq1}) can be obtained by applying the
matching conditions (\ref{eqm}) to appropriate $Q\bar Q$
configurations. To subtract the full QCD amplitude of $Q\bar Q$
state of particular angular momentum, the covariant projection
method is adopted. In practice, the optical theorem can relate the
imaginary part of the QCD amplitude $\mathcal{A}$ in (\ref{eqm}) to
the parton level decay width~\cite{Petrelli,Maltoni}
\begin{equation}
\Gamma(Q\bar Q[n]\rightarrow LFs)=\frac{1}{2M}\langle Q\bar
Q[n]|\mathcal{O}[n]|Q\bar
Q[n]\rangle_{NR}^{LO}\overline{\sum}\int|\mathcal{M}(Q\bar
Q[n]\rightarrow
LFs)|^2\mbox{d}\Phi,\label{Defination:M&Gamma}\end{equation}
where $LFs$ denote the gluons or light quarks and $[n]$ denotes the
configuration of the $Q\bar Q$. The state $|Q\bar Q[n]\rangle$ has
been normalized relativistically as one composite state with mass
$M=2E_Q$, except that in the matrix element in
(\ref{Defination:M&Gamma}), where the state is normalized
non-relativistically to match the results in perturbative NRQCD
conveniently. The super-index $LO$ of the matrix element means that
it is evaluated at tree level, and we always use the abbreviation
$\langle\mathcal{O}[n]\rangle_{LO}$ to represent it in our
calculations. Moreover, the summation/average of the color and
polarization for the final/initial state has been implied by the
symbol $\overline{\sum}$.

For spin-singlet states with $L=0,L=1$ and $L=2$, the amplitudes
$\mathcal{M}$ defined in (\ref{Defination:M&Gamma}) are given
by~\cite{Petrelli}
\begin{eqnarray}\label{M:CovariantProjection}
\mathcal{M}((Q\overline{Q})_{^{1}S_{0}}^{[1,8]}\rightarrow LFs)&=&
\sqrt{\frac{2}{M}}Tr[\mathcal{C}^{[1,8]}\Pi^{0}\mathcal{M}^{am}]|_{q=0},\nonumber\\
\mathcal{M}((Q\overline{Q})_{^{1}P_{1}}^{[8]}\rightarrow
LFs)&=&\epsilon_{\alpha}^{[P]}
\sqrt{\frac{2}{M}}\frac{\textrm{d}}{\textrm{d}q_{\alpha}}Tr[\mathcal{C}^{[8]}\Pi^{0}\mathcal{M}^{am}]|_{q=0},\nonumber\\
\mathcal{M}((Q\overline{Q})_{^{1}D_{2}}^{[1]}\rightarrow
LFs)&=&\frac{1}{2}\epsilon_{\alpha\beta}^{[D]}
\sqrt{\frac{2}{M}}\frac{\textrm{d}^2}{\textrm{d}q_{\alpha}\textrm{d}q_{\beta}}Tr[\mathcal{C}^{[1]}\Pi^{0}\mathcal{M}^{am}]|_{q=0},
\end{eqnarray} where $\mathcal{M}^{am}$ denotes the parton-level
amplitude amputated of the heavy quark spinors, and
$\epsilon_{\alpha}^{[P]}$ and $\epsilon_{\alpha\beta}^{[D]}$ are the
polarization tensors for $L=P,D$ states respectively. The factor
$\sqrt{\frac{2}{M}}=\frac{\sqrt{2M}}{\sqrt{2E_Q}\sqrt{2E_Q}}$ comes
from the normalization of the composite state $|Q\bar Q[n]\rangle$.
For color singlet and octet states, the color projectors are
$\mathcal{C}^{[1]}=\frac{\delta_{ij}}{\sqrt{N_c}}$ and
$\mathcal{C}^{[8]}=\sqrt{2}(T_{a})_{ij}$
respectively~\cite{Petrelli}. The covariant spin-singlet projector
$\Pi^0$ in (\ref{M:CovariantProjection}) is defined by
\begin{equation}\label{defination:Pi0}
\Pi^0=\sum_{s\bar{s}}u(s)\bar{v}(\bar{s})\langle\frac{1}{2},s;\frac{1}{2},\bar{s}|0,0\rangle.
\end{equation}
The explicit form of $\Pi^0$ in D dimensions will be discussed in
the latter subsection.

The sums over polarization tensors for $\epsilon_{\alpha}^{[P]}$ and
$\epsilon_{\alpha\beta}^{[D]}$ in D dimensions are:
\begin{subequations}
\begin{align}
\sum_{J_z}\epsilon^{[P]}_{\alpha}\epsilon^{[P]\ast}_{\alpha^{\prime}}=\Pi_{\alpha
\alpha^{\prime}},
\end{align}
\begin{align}
\sum_{J_z}\epsilon^{[D]}_{\alpha\beta}\epsilon^{[D]\ast}_{\alpha^{\prime}\beta^{\prime}}=
\frac{1}{2}(\Pi_{\alpha\alpha^{\prime}}\Pi_{\beta\beta^{\prime}}
+\Pi_{\alpha\beta^{\prime}}\Pi_{\alpha^{\prime}\beta})-
\frac{1}{D-1}\Pi_{\alpha\beta}\Pi_{\alpha^{\prime}\beta^{\prime}}.
\end{align}
\end{subequations}
Here $\Pi_{\alpha \alpha^{\prime}}$ is defined as
\begin{equation}
\Pi_{\alpha \alpha^{\prime}}=-g_{\alpha
\alpha^{\prime}}+\frac{P_\alpha P_{\alpha^{\prime}}}{M^2},
\end{equation}
where $P$ is the total momentum of $Q\bar Q$, and $P^2=M^2=4E_Q^2$.

Needless to say, the final result should be independent on the
normalization convention of the $Q\bar Q$ state. If one wants to
apply NR normalization thoroughly in the calculations, one needs to
eliminate the factors $1/(2M)$ in (\ref{Defination:M&Gamma}) and
$\sqrt{2/M}$ in (\ref{M:CovariantProjection}), and then to replace
the covariant spinors in (\ref{defination:Pi0}) with the NR ones
with the normalization condition: $u^\dag u=v^\dag v=1$.

\subsection{Discussions on $\gamma^5$ scheme and projection operator}

We will do our calculations in dimensional regularization scheme
both for QCD and NRQCD. Since we are only dealing with the
spin-singlet Fock states, there will be the problem of definition of
$\gamma^5$ in D dimensions. In our calculation, the 't Hooft-Veltman
(HV) scheme\cite{thooft,Petrelli} is introduced:
\begin{eqnarray}
\{\gamma^5,\gamma^\mu\}&=&0,\hspace{0.3cm}\mu=0,1,2,3\nonumber\\
\left[\gamma^5,\gamma^{\mu}\right]&=&0,\hspace{0.3cm}\mu=4,\cdots,D-1.
\label{gamma5}\end{eqnarray} And the $\gamma^5$ matrix can be
represented as~\cite{Novotny}:
\begin{equation}
\gamma^5=-\frac{i}{4!}\epsilon^{\mu\nu\rho\sigma}\gamma_\mu\gamma_\nu\gamma_\rho\gamma_\sigma.
\end{equation}
The calculation involving $\gamma^5$ is carried out in D dimensions,
where $\epsilon^{\mu\nu\rho\sigma}$ and $\gamma^\mu$ are all defined
in D dimensions. Other prescriptions may be found in
literatures~\cite{Mertig,Kreimer,Larin,BM}.

In four dimensions, the covariant spin-singlet projector $\Pi^0$
defined in (\ref{defination:Pi0}) can be given by (see,
e.g.,~\cite{Bodwin02:NRcorections})
\begin{equation}\label{Pi0:form1}
\Pi^0=\frac{1}{2\sqrt{2}(E_Q+m_Q)}(\frac{\slashed{P}}{2}+\slashed{q}+m_Q)
\frac{(\slashed{P}+M)}{M}\gamma^5(\frac{\slashed{P}}{2}-\slashed{q}-m_Q),
\end{equation}
where $q$ is half of the relative momentum of the heavy quark pair.
The form in (\ref{Pi0:form1}) can not keep $\mathbf{C}$ parity
conservation
in D dimensions because $(\slashed{P}+M)\gamma^5$ can not keep an
invariant form under charge conjugation transformation in $D\neq4$
dimensions in the HV scheme, which can be easily seen by applying
(\ref{gamma5}). This problem can be solved by replacing it by the
following two operators. For spin singlet states the spin projectors
of incoming heavy quark pairs at any order in $v^{2}$ are given by
\begin{equation}
\Pi^{0}=\frac{1}{2\sqrt{2}(E_Q+m_Q)}(\frac{\slashed{P}}{2}+\slashed{q}+m_Q)
\frac{[(\slashed{P}+M)\gamma^5+\gamma^5(-\slashed{P}+M)]}{2M}(\frac{\slashed{P}}{2}-\slashed{q}-m_Q)
\end{equation}
from~\cite{Quantum} and
\begin{equation}
\Pi^{0}=\frac{1}{2\sqrt{2}(E_Q+m_Q)}(\frac{\slashed{P}}{2}+\slashed{q}+m_Q)
\frac{(\slashed{P}+M)\gamma^5(-\slashed{P}+M)}{2M^2}(\frac{\slashed{P}}{2}-\slashed{q}-m_Q)
\end{equation}
from~\cite{Keung}. The above two projection operators both give
correct results and keep $\mathbf{C}$ parity conservation.

\section{Full QCD Calculation}
In this section, we calculate the imaginary part of $Q\bar{Q}$
forward scattering amplitude, or equivalently, the parton-level
decay width $\Gamma$ defined in (\ref{Defination:M&Gamma}). In the
calculation, we use {\tt FeynArts}~\cite{FeynArts} to generate the
Feynman diagrams and amplitudes and {\tt FeynCalc}~\cite{FeynCalc}
for the tensor reduction. We regularize the ultraviolet(UV) and
infrared(IR) divergence in dimensional regularization scheme and
extend the covariant projection method into $D=4-2\epsilon$
dimensions as has been mentioned.

The leading order subprocesses in $\alpha_s$ are the annihilations
of $Q\bar Q[n]$ into two gluons, where $n$ can be any configurations
of the Fock states listed in (\ref{FockState}). The Feynman diagrams
at LO of $\alpha_{s}$ are shown in Fig. 1.
\begin{figure}
\begin{center}
\includegraphics[scale=0.8]{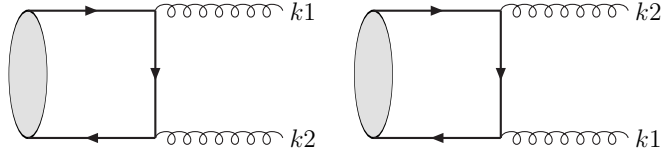}
\caption{Feynman diagrams for ${}^1L_J^{[1,8]}\rightarrow gg$}
\end{center}
\end{figure}
And the results in D dimensions are
\begin{eqnarray}
\Gamma_{\textrm{Born}}(^1S_{0}^{[1]}\rightarrow gg)&=&\frac{C_F\alpha_s^216\pi^2}{m_Q^2}\Phi_2(1-\epsilon)(1-2\epsilon)\langle\mathcal{O}({}^1S_{0}^{[1]})\rangle_{LO},\nonumber\\
\Gamma_{\textrm{Born}}(^1S_{0}^{[8]}\rightarrow gg)&=&\frac{B_F\alpha_s^216\pi^2}{m_Q^2}\Phi_2(1-\epsilon)(1-2\epsilon)\langle\mathcal{O}({}^1S_{0}^{[8]})\rangle_{LO},\nonumber\\
\Gamma_{\textrm{Born}}(^1P_{1}^{[8]}\rightarrow gg)&=&\frac{C_A\alpha_s^24\pi^2}{m_Q^4}\Phi_2\frac{(1-\epsilon)(1-2\epsilon)}{3-2\epsilon}\langle\mathcal{O}({}^1P_{1}^{[8]})\rangle_{LO},\nonumber\\
\Gamma_{\textrm{Born}}(^1D_{2}^{[1]}\rightarrow
gg)&=&\frac{C_F\alpha_s^24\pi^2}{m_Q^6}\Phi_2\frac{(1-2\epsilon)(6\epsilon^2-15\epsilon+8)}{4\epsilon^2-16\epsilon+15}\langle\mathcal{O}({}^1D_{2}^{[1]})\rangle_{LO},
\label{Gamma:Born}\end{eqnarray} where
$B_F=\frac{N_c^2-4}{4N_c}=\frac{5}{12}$ and $\Phi_{(2)}$ is the
two-body phase space in D dimensions:
$\frac{1}{8\pi}\frac{\Gamma(1-\epsilon)}{\Gamma(2-2\epsilon)}(\frac{\pi}{m_{Q}^2})^{\epsilon}$.
The first three results in (\ref{Gamma:Born}) are consistent with
those in Ref.~\cite{Petrelli,Maltoni}. At the Born level, there are
no IR divergences in the results since both the two gluons should be
hard in the rest frame of $Q\bar Q$.

\subsection{Real Corrections}
The real corrections to Born level subprocesses include the decays
into $ggg$ and $q\bar{q}g$ final states. The corresponding Feynman
diagrams are shown in Fig. 2 and Fig. 3. For simplicity, unphysical
polarization summation is used for final state gluons, so diagrams
with ghosts in the final states must be included in calculation when
three gluon vertex appears, in order to cancel the non-physical
degrees of freedom to keep the full results gauge invariant.

\subsubsection{$(Q\bar{Q})_{^1L_{J}^{[1,8]}}\rightarrow ggg $ }

\begin{figure}
\begin{center}
\includegraphics[scale=0.8]{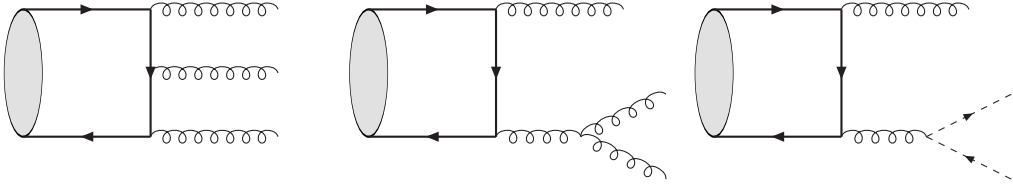}
\caption{Feynman diagrams for ${}^1L_J^{[1,8]}\rightarrow ggg$}
\end{center}
\end{figure}

Our results of S-wave configurations agree with those in
\cite{Petrelli,Maltoni} and are listed below:
\begin{eqnarray}
\Gamma(^1S_{0}^{[1]}\rightarrow
ggg)&=&\frac{C_A\alpha_s}{\pi}\Gamma_{\textrm{Born}}(^1S_0^{[1]}\rightarrow
gg)f_\epsilon(M^2)(\frac{1}{\epsilon^2}+\frac{11}{6\epsilon}+\frac{181}{18}-\frac{23}{24}\pi^2),\nonumber\\
\Gamma(^1S_{0}^{[8]}\rightarrow
ggg)&=&\frac{C_A\alpha_s}{\pi}\Gamma_{\textrm{Born}}(^1S_0^{[8]}\rightarrow
gg)f_\epsilon(M^2)(\frac{1}{\epsilon^2}+\frac{7}{3\epsilon}+\frac{104}{9}-\pi^2),
\label{RC:ggg:Swave}\end{eqnarray}

where
$f_\epsilon(M^2)=(\frac{4\pi\mu^2}{M^2})^\epsilon\Gamma(1+\epsilon)$.
The D-dimension P and D-wave results are:
\begin{eqnarray}
\Gamma(^1P_{1}^{[8]}\rightarrow
ggg)&=&\frac{C_A\alpha_s}{\pi}\Gamma_{\textrm{Born}}(^1P_1^{[8]}\rightarrow
gg)f_\epsilon(M^2)(\frac{1}{\epsilon^2}+\frac{71}{27\epsilon}+\frac{7(-168+25\pi^2)}{162}),\nonumber\\
\Gamma(^1D_{2}^{[1]}\rightarrow
ggg)&=&\frac{C_A\alpha_s}{\pi}\Gamma_{\textrm{Born}}(^1D_2^{[1]}\rightarrow
gg)f_\epsilon(M^2)(\frac{1}{\epsilon^2}+\frac{3}{\epsilon}+\frac{7027}{144}-\frac{277}{64}\pi^2).
\label{RC:ggg:PDwave}\end{eqnarray}

Both soft and collinear IR divergences are there in the results in
(\ref{RC:ggg:Swave}) and (\ref{RC:ggg:PDwave}), and the square pole
$1/\epsilon^2$ comes from the overlap of the soft and the collinear
regions.

\subsubsection{$(Q\bar{Q})_{^1L_{J}^{[1,8]}}\rightarrow q\bar{q}g$}
Another subprocess of light hadronic decay is to $q\bar{q}g$ final
states, and only two graphs make contribution to this subprocess
(shown in Fig.~3).

\begin{figure}
\begin{center}
\includegraphics[scale=0.8]{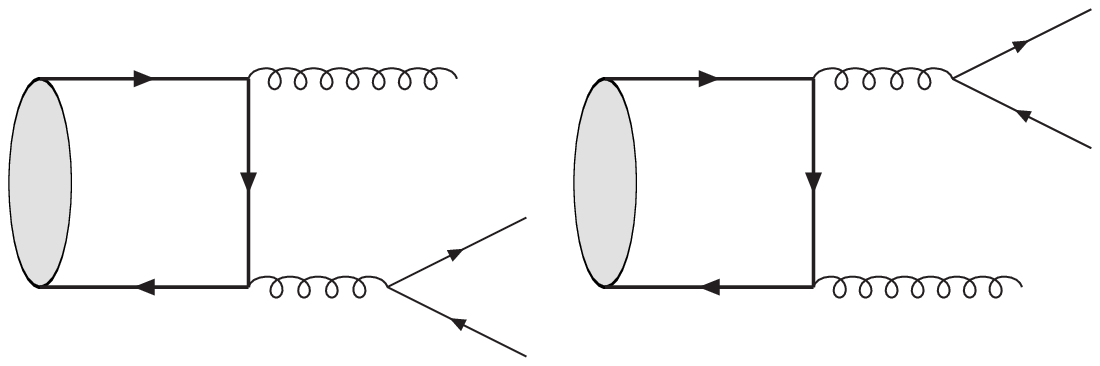}
\caption{Feynman diagrams for ${}^1L_J^{[1,8]}\rightarrow
q\bar{q}g$}
\end{center}
\end{figure}

We get the following results:
\begin{eqnarray}
\Gamma(^1S_{0}^{[1]}\rightarrow
q\bar{q}g)&=&N_f\Gamma_{\textrm{Born}}(^1S_0^{[1]}\rightarrow
gg)\frac{\alpha_s}{\pi}\frac{f_\epsilon(M^2)}{K}T_F(-\frac{2}{3\epsilon}-\frac{16}{9}),\nonumber\\
\Gamma(^1S_{0}^{[8]}\rightarrow
q\bar{q}g)&=&N_f\Gamma_{\textrm{Born}}(^1S_0^{[8]}\rightarrow
gg)\frac{\alpha_s}{\pi}\frac{f_\epsilon(M^2)}{K}T_F(-\frac{2}{3\epsilon}-\frac{16}{9}),\nonumber\\
\Gamma(^1P_{1}^{[8]}\rightarrow
q\bar{q}g)&=&N_f\Gamma_{\textrm{Born}}(^1P_1^{[8]}\rightarrow
gg)\frac{\alpha_s}{\pi}\frac{f_\epsilon(M^2)}{K}T_F(-\frac{2}{3\epsilon}-\frac{16}{9}),\nonumber\\
\Gamma(^1D_{2}^{[1]}\rightarrow
q\bar{q}g)&=&N_f\Gamma_{\textrm{Born}}(^1D_2^{[1]}\rightarrow
gg)\frac{\alpha_s}{\pi}\frac{f_\epsilon(M^2)}{K}T_F(-\frac{2}{3\epsilon}-\frac{16}{9}),
\label{RC:gqq}\end{eqnarray} where $N_f$ is the number of light
flavor quarks. $N_f=3$ and 4 for charmonium and bottomonium
respectively. $T_F=\frac{1}{2}$,
$K=\Gamma(1+\epsilon)\Gamma(1-\epsilon)\simeq1+\epsilon^2\frac{\pi^2}{6}$
and the S-wave results agree with \cite{Petrelli,Maltoni}.

There are only single poles of $\epsilon$ in the results in
(\ref{RC:gqq}) and they can be identified as collinear ones. The
absence of the soft IR divergence can be seen from the diagrams in
Fig.~3. When the momentum of the real gluon goes to zero, it will
decouple from the quark line as an eikonal factor~\cite{Maltoni},
then the results will be zero since $Q\bar Q$ in spin-singlet can
not couple to one virtual gluon.

As will be seen later, the collinear divergences and partial soft IR
ones in (\ref{RC:ggg:Swave}), (\ref{RC:ggg:PDwave}) and
(\ref{RC:gqq}) are canceled by the virtual corrections to the Born
level decay width. The remaining soft IR divergences are those in
$\Gamma(^1P_{1}^{[8]}\rightarrow ggg)$ and
$\Gamma(^1D_{2}^{[1]}\rightarrow ggg)$ from the first diagram in
Fig.~2, which will be absorbed in the renormalization of the
operators $\mathcal{O}_{1,8}(^1S_{0})$ and
$\mathcal{O}_{8}(^1P_{1})$ in perturbative NRQCD. These are just the
general results of the so-called topological factorization discussed
in \cite{BBL}.

\subsection{Virtual Corrections}
There are 23 virtual correction diagrams, including counter-term
diagrams, divided into 9 groups. Representative Feynman diagrams of
each class are shown in Fig. 4. And the others can be found through
reversing the arrows on the quark lines or exchanging the final
state gluons. UV divergences are removed by renormalization. The
definitions of the renormalization constant of QCD gauge coupling
constant $g_{s}=\sqrt{4\pi\alpha_{s}}$, heavy quark mass $m_Q$,
heavy quark field $\psi_Q$, light quark field $\psi_q$ and gluon
field $A_{\mu}$ are:
\begin{equation}
g_{s}^{0}=Z_g g_s,\quad m_{Q}^{0}=Z_{m_Q}m_{Q},\quad
\psi_{Q}^{0}=\sqrt{Z_{2Q}}\psi_{Q},\quad
\psi_{q}^{0}=\sqrt{Z_{2q}}\psi_{q},\quad
A_{\mu}^{0}=\sqrt{Z_3}A_{\mu},
\end{equation}
where the superscript 0 labels bare quantities, and $Z_i=1+\delta
Z_i$. The renormalized constant $Z_{g}$ is defined by
minimal-subtraction ($\overline{MS}$) scheme, and the others by the
on-mass-shell ($OS$) scheme, similar to that in \cite{Klasen}. Then
the results are:
\begin{eqnarray}
\delta Z_{2Q}^{OS}&=&-C_F\frac{\alpha_s
}{4\pi}f_\epsilon(M^2)(\frac{1}{\epsilon_{UV}}+\frac{2}{\epsilon}+6\ln(2)+4),\nonumber\\
\delta Z_{2q}^{OS}&=&-C_F\frac{\alpha_s
}{4\pi}f_\epsilon(M^2)(\frac{1}{\epsilon_{UV}}-\frac{1}{\epsilon}),\nonumber\\
\delta
Z_{3}^{OS}&=&(b_0-C_A)\frac{\alpha_s}{4\pi}f_\epsilon(M^2)(\frac{2}{\epsilon_{UV}}-\frac{2}{\epsilon}),\nonumber\\
\delta
Z_{m_{Q}}^{OS}&=&-3C_F\frac{\alpha_s}{4\pi}f_\epsilon(M^2)(\frac{1}{\epsilon_{UV}}+2\ln(2)+\frac{4}{3}),\nonumber\\
\delta
Z_{g}^{\overline{MS}}&=&-b_0\frac{\alpha_s}{4\pi}f_\epsilon(M^2)(\frac{1}{\epsilon_{UV}}-\ln(\frac{\mu^2}{4m_Q^2})),
\end{eqnarray}
where $b_0=\frac{11C_{A}}{6}-\frac{N_{f}}{3}$.

\begin{figure}
\begin{center}
\includegraphics[scale=0.65]{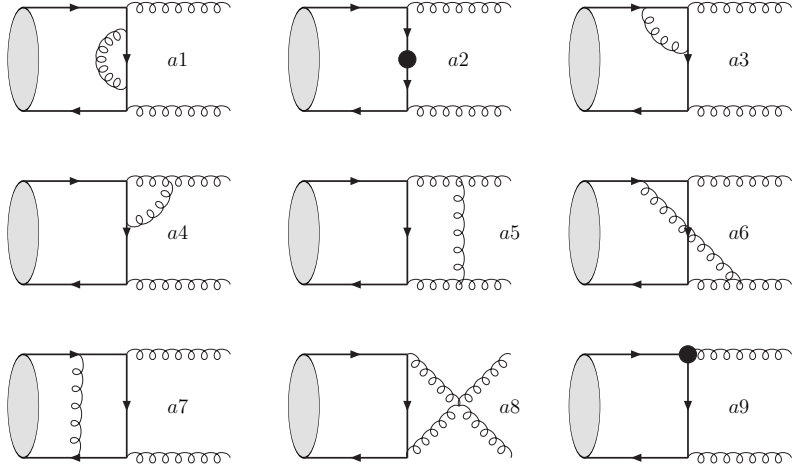}
\caption{One-loop Feynman diagrams for
$(Q\bar{Q})_{^{1}L_{J}^{[1,8]}}\rightarrow gg$ }
\end{center}
\end{figure}

We calculate diagrams one by one and summarize the results in the
following form:
\begin{equation}
\Gamma(^1L_{J}^{[1,8]}\rightarrow gg)_{_{VC}}=
\Gamma(^1L_{J}^{[1,8]}\rightarrow
gg)_{\textrm{Born}}\frac{\alpha_{s}}{\pi}f_{\epsilon}(M^2)
\sum_{k}\mathcal{D}_{k},
\end{equation}
where the results of $\mathcal{D}_{k}$  are listed in Table I-IV. We
add the counter-term diagrams with the corresponding self-energy and
vertex diagrams to show the explicit cancelation of the UV
divergence. There are still IR divergences left, which will be
canceled by those in the real corrections as we have mentioned.
There are also the well-known Coulomb singularities, which have been
regularized by the relative velocity $v$, in Table I-IV. These
singularities can be absorbed by the corresponding matrix element
through the matching condition (\ref{eqm}).

\begin{center}
\begin{table}
\caption{Virtual corrections to
$(Q\bar{Q})_{^1S_{0}^{[1]}}\rightarrow gg$.}
\begin{tabular}{|c|c|c|c|}
     \hline
     Diag.& $\mathcal{D}_{k}$\\
     \hline
       a1+a2 & $C_F(\frac{1}{\epsilon}+1+6\ln2)$ \\
     \hline
     a3+a4+a9  & $-\frac{C_A}{2\epsilon^2}+\frac{1}{\epsilon}(-2C_F-b_0+\frac{C_A}{2})+b_0\ln\frac{\mu^2}{4m_Q^2}
     +C_F(-8\ln2-4+\frac{\pi^2}{4})-\frac{C_A}{2}(-4+\frac{\pi^2}{12})$ \\
     \hline
     a5  & $C_A(-\frac{1}{\epsilon^2}-\frac{1}{\epsilon}-2+2\ln2+\frac{2}{3}\pi^2)$ \\
     \hline
     a6  & $\frac{1}{2}C_A(\frac{1}{\epsilon^2}+\frac{1}{\epsilon}+2-4\ln2-\frac{5}{12}\pi^2)$ \\
      \hline
     a7  & $C_F(\frac{\pi^2}{2v}+\frac{1}{\epsilon}-2+2\ln2)$\\
      \hline
     a8  & $0$ \\
      \hline
\end{tabular}
\end{table}
\end{center}

\begin{center}
\begin{table}
\caption{Virtual corrections to
$(Q\bar{Q})_{^1S_{0}^{[8]}}\rightarrow gg$.}
\begin{tabular}{|c|c|c|c|}
     \hline
     Diag.& $\mathcal{D}_{k}$ \\
     \hline
       a1+a2 & $C_F(\frac{1}{\epsilon}+1+6\ln2)$ \\
     \hline
     a3+a4+a9  & $-\frac{C_A}{2\epsilon^2}+\frac{1}{\epsilon}(-2C_F-b_0+\frac{C_A}{2})+b_0\ln\frac{\mu^2}{4m_Q^2}
     +C_F(-8\ln2-4+\frac{\pi^2}{4})-\frac{C_A}{2}(-4+\frac{\pi^2}{12})$\\
     \hline
     a5  &  $\frac{1}{2}C_A(-\frac{1}{\epsilon^2}-\frac{1}{\epsilon}-2+2\ln2+\frac{2}{3}\pi^2)$\\
     \hline
     a6  &  $0$\\
      \hline
     a7  &  $(C_F-\frac{1}{2}C_A)(\frac{\pi^2}{2v}+\frac{1}{\epsilon}-2+2\ln2)$\\
      \hline
     a8  &  $0$\\
      \hline
\end{tabular}
\end{table}
\end{center}

\begin{center}
\begin{table}
\caption{Virtual corrections to
$(Q\bar{Q})_{^1P_{1}^{[8]}}\rightarrow gg$.}
\begin{tabular}{|c|c|c|}
     \hline
     Diag.& $\mathcal{D}_{k}$\\
     \hline
       a1+a2 & $C_F(\frac{1}{\epsilon}-3+10\ln2)$ \\
     \hline
     a3+a4+a9 & $-\frac{C_A}{2\epsilon^2}-\frac{1}{\epsilon}(2C_F+b_0+\frac{C_A}{2})+b_0\ln\frac{\mu^2}{4m_Q^2}
     +C_F(\frac{\pi^2}{2}-16\ln2)-\frac{C_A}{2}(-6+\frac{\pi^2}{3}+2\ln2)$  \\
     \hline
     a5  &   $\frac{1}{2}C_A(-\frac{1}{\epsilon^2}-\frac{1}{\epsilon}-5+\frac{2}{3}\pi^2+5\ln2)$\\
     \hline
     a6  &  $0$\\
      \hline
     a7  &  $(C_F-\frac{1}{2}C_A)(\frac{\pi^2}{2v}+\frac{1}{\epsilon}-2+2\ln2)$\\
      \hline
     a8  &  $C_A(-\frac{1}{2}+\frac{\ln2}{2})$\\
      \hline
\end{tabular}
\end{table}
\end{center}

\begin{center}
\begin{table}
\caption{Virtual corrections to
$(Q\bar{Q})_{^1D_{2}^{[1]}}\rightarrow gg$.}
\begin{tabular}{|c|c|c|}
     \hline
     Diag.& $\mathcal{D}_{k}$\\
     \hline
       a1+a2 &$C_F(\frac{1}{\epsilon}-10+22\ln2)$  \\
     \hline
     a3+a4+a9  & $-\frac{C_A}{2\epsilon^2}-\frac{1}{\epsilon}(2C_F+b_0+C_A)+b_0\ln\frac{\mu^2}{4m_Q^2}
     +C_F(10+\frac{3}{4}\pi^2-38\ln2)-\frac{C_A}{2}(-\frac{15}{8}+\frac{7}{12}\pi^2+2\ln2)$ \\
     \hline
     a5  &  $C_A(-\frac{1}{\epsilon^2}-\frac{1}{\epsilon}-8+\frac{2}{3}\pi^2+12\ln2)$\\
     \hline
     a6  &  $\frac{1}{2}C_A(\frac{1}{\epsilon^2}+\frac{69}{8}-\frac{7}{6}\pi^2-16\ln2)$\\
      \hline
     a7  &   $C_F(\frac{\pi^2}{2v}+\frac{1}{\epsilon}-2+4\ln2)$\\
      \hline
     a8  &  $0$\\
      \hline
\end{tabular}
\end{table}
\end{center}

\subsection{Summary of the QCD results}
Combining the real and virtual correction results together and
translating the parton-level decay width back to the imaginary part
of the forward scattering amplitude, we get the full QCD results up
to $\mathcal{O}(\alpha_s^3)$:
\begin{subequations}\label{FullQCDRe}
\begin{align}
(2\textrm{Im}\mathcal{A}(Q\bar Q[^1S_{0}^{[1]}]\to Q\bar
Q[^1S_{0}^{[1]}]))\Big{|}_{\textrm{pert
QCD}}=\{\frac{8\pi\alpha_s^2}{3m_Q^2}(1+\frac{2\pi\alpha_s}{3v})+\frac{\alpha_s^3}{27m_Q^2}[4(477-16N_f)\nonumber\\
+12(33-2N_f)\ln\frac{\mu^2}{4m_Q^2}-93\pi^2]\}\langle\mathcal{O}({}^1S_0^{[1]})\rangle_{LO},\nonumber\\
\end{align}
\begin{align}
(2\textrm{Im}\mathcal{A}(Q\bar Q[^1S_{0}^{[8]}]\to Q\bar
Q[^1S_{0}^{[8]}]))\Big{|}_{\textrm{pert
QCD}}=\{\frac{5\pi\alpha_s^2}{6m_Q^2}(1-\frac{\pi\alpha_s}{12v})+\frac{5\alpha_s^3}{432m_Q^2}[16(153-4N_f)\nonumber\\
+12(33-2N_f)\ln\frac{\mu^2}{4m_Q^2}-129\pi^2]\}\langle\mathcal{O}({}^1S_0^{[8]})\rangle_{LO},\nonumber\\
\end{align}
\begin{align}
(2\textrm{Im}\mathcal{A}(Q\bar Q[^1P_{1}^{[8]}]\to Q\bar
Q[^1P_{1}^{[8]}]))\Big{|}_{\textrm{pert
QCD}}=\{\frac{\pi\alpha_s^2}{2m_Q^4}(1-\frac{\pi\alpha_s}{12v})-\frac{19\alpha_s^3}{18m_Q^4}[\frac{1}{\epsilon}-\gamma_E+\ln(4\pi)]\nonumber\\
+\frac{\alpha_s^3[2(3(-8N_f-21\ln(2)-229)+119\pi^2)
-3(6N_f-61)\ln(\frac{\mu^2}{4m_Q^2})]}{108m_Q^4}\}\langle\mathcal{O}({}^1P_1^{[8]})\rangle_{LO},\nonumber\\
\end{align}
\begin{align}
(2\textrm{Im}\mathcal{A}(Q\bar Q[^1D_{2}^{[1]}]\to Q\bar
Q[^1D_{2}^{[1]}]))\Big{|}_{\textrm{pert
QCD}}=\{\frac{16\pi\alpha_s^2}{45m_Q^6}(1+\frac{2\pi\alpha_s}{3v})-\frac{8\alpha_s^3}{9m_Q^6}[\frac{1}{\epsilon}-\gamma_E+\ln(4\pi)]\nonumber\\
-\frac{\alpha_s^3[4(128N_f+1008\ln(2)-19509)
+192(N_f-9)\ln(\frac{\mu^2}{4m_Q^2})+7263\pi^2]}{1620m_Q^6}\}\langle\mathcal{O}({}^1D_2^{[1]})\rangle_{LO},
\end{align}
\end{subequations}
where the states $|Q\bar Q[n]\rangle$ in the l.h.s. and the r.h.s.
should be understood to have been normalized under the same
condition. Moreover, the equalities in (\ref{FullQCDRe}) are
independent on the normalization conventions. That is, the state
$|Q\bar Q[n]\rangle$ can be normalized either relativistically or
non-relativistically, either as a composite state or as a discrete
state. Therefore it is convenient to use $\textrm{Im}\mathcal{A}$ to
do the matching calculations, and we will use the abbreviation
$\textrm{Im}\mathcal{A}(n)$ to represent the amplitudes in
(\ref{FullQCDRe}).

The remaining infrared divergences and Coulomb singularities in
(\ref{FullQCDRe}) will be precisely repeated in the radiative
corrections of the matrix elements in perturbative NRQCD in next
section. The finite short distance coefficients will be obtained
after matching calculations.

\section{NRQCD Results and Operator Evolution Equations}

In this section, we calculate the NRQCD corrections to the four
fermion operators in D dimensions. As we have mentioned, we will
adopt the method of regions~\cite{Beneke} to avoid the mismatch of
the loop momenta in different regions. Furthermore, each loop
integral in this method contributes only to a single power in $v$,
thus one can do the power counting before the integral has been
explicitly done.

Since the ultrasoft and the QCD scale are comparable,
$m_Q\,v^2\sim\Lambda_{QCD}$, for both charmonium and bottomonium,
there are only two low-energy scales to be considered in NRQCD,
which satisfy the inequality $m_Q\,v\gg m_Q\,v^2$. Thus, the
nontrivial contributions to the NRQCD loop integrals come only from
the following three regions:
\begin{equation}
\begin{array}{l}
{\textrm{soft}: \qquad\; A^{\mu}_{s}:\quad k_{0}\sim |\vec{k}|\sim
m_{Q}v,\qquad \quad\Psi_{s}:T\sim
|\vec{p}|\sim m_{Q}v},\\
{\textrm{potential}: A^{\mu}_{p}:\quad k_{0}\sim m_{Q}v^{2},
|\vec{k}|\sim m_{Q}v,\; \Psi_{p}:T\sim m_{Q}v^{2},\;
|\vec{p}|\sim m_{Q}v},\\
{\textrm{ultrasoft}\,: A^{\mu}_{u}:\quad k_{0}\sim |\vec{k}|\sim
m_{Q}v^{2},}
\end{array}
\label{defination:regions}\end{equation} where $k_\nu$ and $p_\nu$
are the momenta of gluon field and heavy quark field respectively
and $T=p_0-m_Q$. The loop momenta running in these regions scale as
those of the corresponding gluons in dimensional regularization
scheme. One can check that the other regions, such as that with
$k_0\sim m_Q\,v$ and $\vec{k}\sim m_Q\,v^2$, have no contributions
to the NRQCD loop integrals in dimensional regularization scheme. In
the view point of effective field theory, the five modes defined in
(\ref{defination:regions}) can be all treated as the effective
fields in NRQCD. Only after parting these low-energy modes
sufficiently like what has been done in (\ref{defination:regions}),
the homogeneous power counting rules can be gotten.

In practice, we use the NRQCD Feynman rules~\cite{Griesshammer}
derived in Coulomb gauge for the three regions (or the five
low-energy modes) in our calculations. These rules are shown in Fig.
5 and 6, where
$\delta_{tr}^{ij}=\delta^{ij}-\frac{k^{i}k^{j}}{|\mathbf{k}|^{2}}$.
The Feynman rules for anti-heavy quark could be obtained by charge
conjugation symmetry.
\begin{figure}
\begin{center}
\includegraphics[scale=0.65]{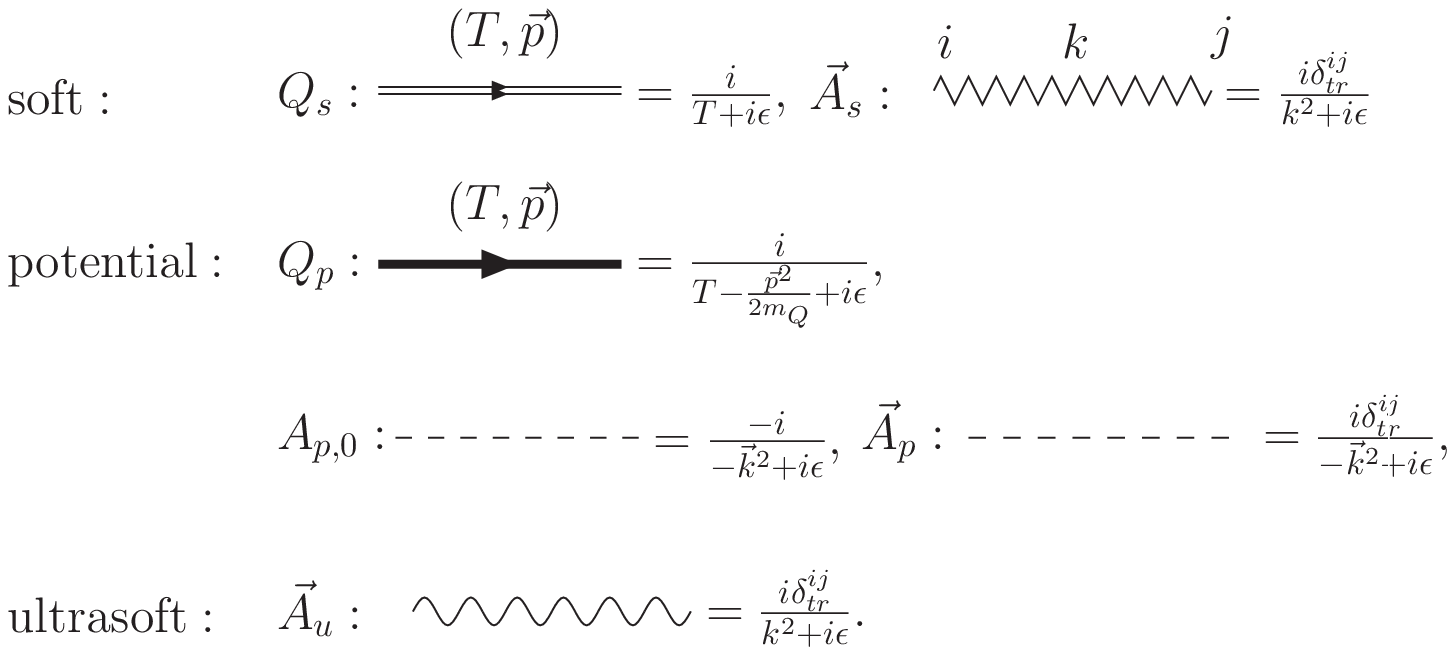}
\caption{NRQCD Feynman rules for heavy quark and gluon propagators
in different regions }
\end{center}
\end{figure}

\begin{figure}
\begin{center}
\includegraphics[scale=0.65]{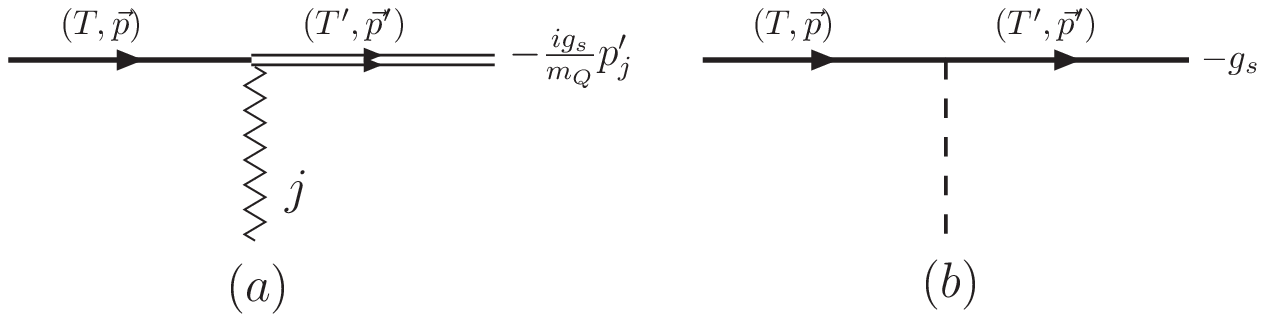}
\caption{NRQCD Feynman rules for heavy quark and gluon vertices }
\end{center}
\end{figure}

The Coulomb singularities calculated in full QCD theory correspond
to the potential region, while the soft divergences to the soft one.
The LO Feynman diagrams for matrix elements are shown in Fig.~7. The
onshell external quark lines lie in potential region. At NLO in
$\alpha_s$, only six classes of Feynman diagrams shown in Fig.~8
need to be calculated for our purpose~\cite{He}. The first four
diagrams (a)-(d) have inner gluon lines connecting with one incoming
quark line and one outgoing quark line, and the soft region will
give the lowest order nontrivial result in $v$~\cite{He}. In the
last two diagrams (e) and (f) the inner gluon line joints two
incoming or outgoing quark lines, and only the potential region has
non-vanishing real value~\cite{He}. The self-energy diagrams in the
external legs are dropped in accordance with the on-shell
renormalization scheme used in the full QCD calculation.

\begin{figure}
\begin{center}
\includegraphics[scale=0.5]{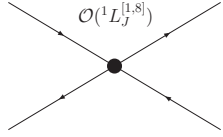}
\caption{NRQCD Feynman diagrams for LO Matrix Elements}
\end{center}
\end{figure}

\begin{figure}
\begin{center}
\includegraphics[scale=0.5]{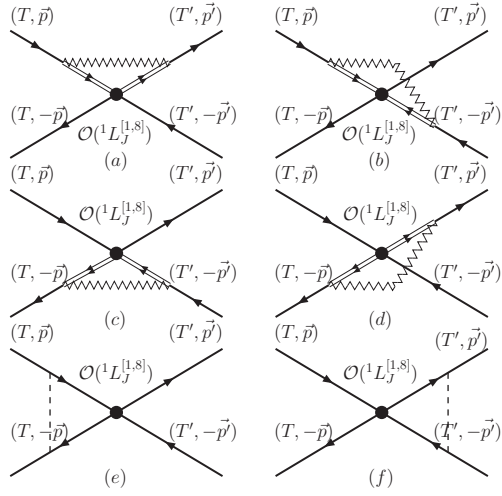}
\caption{NRQCD Feynman diagrams for NLO Matrix Elements }
\end{center}
\end{figure}

We present here the detailed calculation of the NLO correction to
the P-wave octet operator $\mathcal{O}({}^1P_1^{[8]})$. The LO
result $\langle \mathcal{O}({}^1P_1^{[8]})\rangle_{\textrm{Born}}$
is trivial. Using the Feynman rules for propagators of heavy quark
and gluon in soft region and for the heavy quark gluon vertex
between potential and soft regions, the loop integral of diagram (a)
is:
\begin{equation}
I_{a}=\frac{ig_{s}^{2}}{m_{Q}^{2}}\int \frac{d^{D}k}{(2\pi)^{D}}
\frac{\mathbf{p}\cdot\mathbf{p'}-(\mathbf{p}\cdot\mathbf{k})(\mathbf{p'}\cdot\mathbf{k})/\mathbf{k}^{2}}
{k_{0}^{2}-\mathbf{k}^{2}+i\epsilon}\frac{1}{k_0-i\epsilon}\frac{1}{k_0-i\epsilon}.
\end{equation}
After performing the contour-integration of
$k_{0}=|\mathbf{k}|-i\epsilon$, we get\footnote{Since the quark
propagator poles should be taken into account in the potential
region, one only needs to evaluate the contribution from the gluon
pole in the soft region~\cite{Beneke}.}
\begin{equation}\label{integral:soft}
I_{a}=\frac{g_{s}^{2}}{2m_{Q}^{2}}\int \frac{d^{D-1}k}{(2\pi)^{D-1}}
\frac{\mathbf{p}\cdot\mathbf{p'}-(\mathbf{p}\cdot\mathbf{k})(\mathbf{p'}\cdot\mathbf{k})/\mathbf{k}^{2}}
{|\mathbf{k}|^{3}},
\end{equation}
which is both infrared and ultraviolet divergent. The integral in
(\ref{integral:soft}) is scaleless, so vanishes in dimensional
regularization. That is, the UV pole will be canceled by the IR one.
But the result is nontrivial:
\begin{equation}
I_{a}=\frac{\alpha_{s}}{3\pi m_{Q}^2}
(\frac{1}{\epsilon_{UV}}-\frac{1}{\epsilon})\mathbf{p}\cdot\mathbf{p'}.
\end{equation}
The integrals of (b)-(d) in Fig. 8 could be evaluated in the same
way, and their results are:
\begin{equation}
I_{b-d}=\frac{\alpha_{s}}{3\pi m_{Q}^2}
(\frac{1}{\epsilon_{UV}}-\frac{1}{\epsilon})\mathbf{p}\cdot\mathbf{p'}.
\end{equation}

Making use of the Feynman rules for heavy quarks and gluon in
potential region, we obtain the loop integral of diagram (e):
\begin{equation}
I_{e}=-ig_{s}^{2}\int \frac{d^D
k}{(2\pi)^{D}}\frac{1}{\mathbf{k}^{2}}
\frac{1}{T+k_{0}-\frac{(\mathbf{p}+\mathbf{k})^2}{2m_{Q}}+i\epsilon}\;
\frac{1}{T-k_{0}-\frac{(\mathbf{p}+\mathbf{k})^2}{2m_{Q}}+i\epsilon},
\end{equation}
where $T=\frac{|\mathbf{p}|^{2}}{2m_{Q}}$. Integrating $k_0$, we
have
\begin{equation}
I_{e}=g_{s}^{2}m_{Q}\int
\frac{d^{D-1}k}{(2\pi)^{D-1}}\frac{1}{\mathbf{k}^{2}}
\frac{1}{\mathbf{k}^{2}+2\mathbf{p}\cdot\mathbf{k}-i\epsilon}.
\end{equation}
This integral could be performed directly by introducing
$v=\frac{|\mathbf{p}|}{m_{Q}}$, and we get the Coulomb singularity:
\begin{equation}
I_{e}=\frac{\alpha_{s}\pi}{4v}(1-\frac{i}{\pi}(\frac{1}{\epsilon}-\ln(\frac{m_{Q}^{2}v^{2}}{\pi\mu^{2}})-\gamma_
E)).
\end{equation}
The integral of diagram (f) gives the same Coulomb singularity but
with opposite sign in the imaginary part.

The color structures of diagrams (a,c), (b,d) and (e,f) are obtained
by color decomposition and are listed below respectively:
\begin{eqnarray}
\sqrt{2}T^{a}T^{b}\otimes
T^{b}\sqrt{2}T^{a}&=&C_{F}\frac{1}{\sqrt{3}}\otimes
\frac{1}{\sqrt{3}}+\frac{N_{c}^{2}-2}{2N_{c}} \sqrt{2}T^{c}\otimes
\sqrt{2}T^{c},\nonumber\\
\sqrt{2}T^{a}T^{b}\otimes
\sqrt{2}T^{a}T^{b}&=&C_{F}\frac{1}{\sqrt{3}}\otimes
\frac{1}{\sqrt{3}}+\frac{-2}{2N_{c}} \sqrt{2}T^{c}\otimes
\sqrt{2}T^{c},\nonumber\\
T^{b}\sqrt{2}T^{a}T^{b}\otimes\sqrt{2}T^{a}
&=&(C_{F}-\frac{1}{2}C_{A})\sqrt{2}T^{c}\otimes\sqrt{2}T^{c}.
\end{eqnarray}

Summing over each integral multiplied by the according color factor
we get NRQCD matrix element of P-wave operator at NLO, which is UV
divergent and needs to be renormalized:
\begin{equation}
\begin{array}{c}
\langle{\mathcal{O}^{0}(^1P_{1}^{[8]})\rangle_{NLO}=
\{(1+\frac{\alpha_{s}\pi}{2v}(C_{F}-\frac{1}{2}C_{A}))\sqrt{2}T^{c}\otimes\sqrt{2}T^{c}+}\\
{\frac{4\alpha_{s}(\frac{\mu}{\mu_{\Lambda}})^{2\epsilon}}{3\pi
m_{Q}^{2}}(\frac{1}{\epsilon_{_{UV}}}-\frac{1}{\epsilon})
[C_{F}\frac{1}{\sqrt{3}}\otimes\frac{1}{\sqrt{3}}
+B_{F}\sqrt{2}T^{c}\otimes
\sqrt{2}T^{c}]\mathbf{p}\cdot\mathbf{p'}\}
\langle\mathcal{\bar{O}}(^1P_{1})}\rangle_{LO},
\end{array}
\end{equation}
where we have used
$\langle\mathcal{O}^0(^1P_{1}^{[8]})\rangle=\langle\mathcal{\bar{O}}^0(^1P_{1})\rangle
\sqrt{2}T^{a}\otimes\sqrt{2}T^{a}$ and the superscript 0 means the
bare operator. Before doing the operator renormalization, we first
re-express the bare result as
\begin{equation}
\begin{array}{c}
{\langle\mathcal{O}^{0}(^1P_{1}^{[8]})\rangle_{NLO}=
(1+\frac{\alpha_{s}\pi}{2v}(C_{F}-\frac{1}{2}C_{A}))\langle\mathcal{O}(^1P_{1}^{[8]})\rangle_{LO}}
\\
{+\frac{4\alpha_{s}(\frac{\mu}{\mu_{\Lambda}})^{2\epsilon}}{3\pi
m_{Q}^{2}}(\frac{1}{\epsilon_{_{UV}}}-\frac{1}{\epsilon})(
C_{F}\langle\mathcal{O}(^1D_{2}^{[1]})\rangle_{LO}
+B_{F}\langle\mathcal{O}(^1D_{2}^{[8]})\rangle_{LO})}.
\end{array}
\end{equation}
From the above equation we can see that the color-octet P-wave
operator is mixed with the color-singlet D-wave operator at NLO in
$\alpha_s$. We define the renormalized operator
$\mathcal{O}^{R}(^1P_{1}^{[8]})$ through~\cite{Klasen}
\begin{equation}
\begin{array}{c}
\langle\mathcal{O}^{0}(^1P_{1}^{[8]})\rangle_{NLO}=\langle\mathcal{O}^{R}(^1P_{1}^{[8]})\rangle_{\textrm{NLO}}
+\frac{4\alpha_{s}(\frac{\mu}{\mu_{\Lambda}})^{2\epsilon}}{3\pi
m_{Q}^{2}}(\frac{1}{\epsilon_{_{UV}}}+\ln4\pi-\gamma_{E})(C_{F}\langle\mathcal{O}(^1D_{2}^{[1]})\rangle_{LO}
\\
+B_{F}\langle\mathcal{O}(^1D_{2}^{[8]})\rangle_{LO}).
\end{array}
\end{equation}
Here the $\overline{MS}$ renormalization scheme is adopted. The
matrix element of the renormalized operator is UV finite, but still
has IR divergence term which will cancel the infrared divergent
D-wave full QCD result. And it also has the Coulomb singularity,
which is the same as that appearing in the full QCD virtual
correction:
\begin{equation}\label{P8}
\begin{array}{c}
\langle{\mathcal{O}^{R}(^1P_{1}^{[8]})\rangle_{NLO}=
(1+\frac{\alpha_{s}\pi}{2v}(C_{F}-\frac{1}{2}C_{A}))\langle\mathcal{O}(^1P_{1}^{[8]})\rangle_{LO}}
\\
{+\frac{4\alpha_{s}(\frac{\mu}{\mu_{\Lambda}})^{2\epsilon}}{3\pi
m_{Q}^{2}}(-\frac{1}{\epsilon}-\ln 4\pi+\gamma_{E})(
C_{F}\langle\mathcal{O}(^1D_{2}^{[1]})\rangle_{LO}
+B_{F}\langle\mathcal{O}(^1D_{2}^{[8]})\rangle_{LO})}.
\end{array}
\end{equation}
Here, the matrix element of $\mathcal{O}(^1D_{2}^{[8]})$ is at
higher order in $v$ in our case and therefore can be eliminated. The
matrix elements of S-wave singlet and octet operators and that of
the D-wave singlet operator could be computed in the same way:
\begin{equation}\label{S1}
\begin{array}{c}
{\langle\mathcal{O}^{R}(^1S_{0}^{[1]})\rangle_{NLO}=
(1+\frac{\alpha_{s}\pi}{2v}C_{F})\langle\mathcal{O}(^1S_{0}^{[1]})\rangle_{LO}}
\\
{+\frac{1}{2N_c}\frac{4\alpha_{s}(\frac{\mu}{\mu_{\Lambda}})^{2\epsilon}}{3\pi
m_{Q}^{2}}(-\frac{1}{\epsilon}-\ln 4\pi+\gamma_{E})
\langle\mathcal{O}(^1P_{1}^{[8]})\rangle_{LO}},
\end{array}
\end{equation}
\begin{equation}\label{S8}
\begin{array}{c}
{\langle\mathcal{O}^{R}(^1S_{0}^{[8]})\rangle_{NLO}=
(1+\frac{\alpha_{s}\pi}{2v}(C_{F}-\frac{1}{2}C_{A}))\langle\mathcal{O}(^1S_{0}^{[8]})\rangle_{LO}}
\\
{+B_{F}\frac{4\alpha_{s}(\frac{\mu}{\mu_{\Lambda}})^{2\epsilon}}{3\pi
m_{Q}^{2}}(-\frac{1}{\epsilon}-\ln 4\pi+\gamma_{E})
\langle\mathcal{O}(^1P_{1}^{[8]}\rangle_{LO}}+...\,,
\end{array}
\end{equation}
\begin{equation}{\langle\mathcal{O}^{R}(^1D_{2}^{[1]})\rangle_{NLO}=
(1+\frac{\alpha_{s}\pi}{2v}C_{F})\langle\mathcal{O}(^1D_{2}^{[1]})\rangle_{LO}}+...\,,
\end{equation}
where "$...$" denotes terms at higher order in $v$.

Finally combining the matrix elements given above with the short
distance coefficients accordingly, we get the forward scattering
amplitudes for $^1L_J$ states computed by NRQCD effective theory,
which are summarized below:
\begin{subequations}\label{NRQCD}
\begin{align}
(2\textrm{Im}\mathcal{A}(^{1} S_{0}^{[1]}))\Big{|}_{\textrm{pert
NRQCD}}=\frac{2\textrm{Im}f(^1S_0^{[1]})}{m_Q^2}(1+C_F\frac{\alpha_s\pi}{2\upsilon})
\langle\mathcal{O}(^1S_{0}^{[1]})\rangle_{LO},
\end{align}
\begin{align}
(2\textrm{Im}\mathcal{A}(^{1} S_{0}^{[8]}))\Big{|}_{\textrm{pert
NRQCD}}=\frac{2\textrm{Im}f(^1S_0^{[8]})}{m_Q^2}[1+(C_F-\frac{1}{2}C_A)\frac{\alpha_s\pi}{2\upsilon}]
\langle\mathcal{O}(^1S_{0}^{[8]})\rangle_{LO},
\end{align}
\begin{align}
(2\textrm{Im}\mathcal{A}(^{1} P_{1}^{[8]}))\Big{|}_{\textrm{pert
NRQCD}}=\{\frac{2\textrm{Im}f(^1P_1^{[8]})}{m_Q^4}[1+(C_F-\frac{1}{2}C_A)\frac{\alpha_s\pi}{2\upsilon}]
\nonumber\\-\frac{1}{2N_c}\frac{4\alpha_s}{3\pi
m_Q^4}\frac{2\textrm{Im}f(^1S_0^{[1]})}{\epsilon}-\frac{4\alpha_sB_F}{3\pi
m_Q^4}\frac{2\textrm{Im}f(^1S_0^{[8]})}{\epsilon}\}
\langle\mathcal{O}(^1P_{1}^{[8]})\rangle_{LO},
\end{align}
\begin{align}
(2\textrm{Im}\mathcal{A}(^{1} D_{2}^{[1]}))\Big{|}_{\textrm{pert
NRQCD}}=[\frac{2\textrm{Im}f(^1D_2^{[1]})}{m_Q^6}(1+C_F\frac{\alpha_s\pi}{2\upsilon})
\nonumber\\-\frac{4\alpha_sC_F}{3\pi
m_Q^6}\frac{2\textrm{Im}f(^1P_1^{[8]})}{\epsilon}]
\langle\mathcal{O}(^1D_{2}^{[1]})\rangle_{LO}.
\end{align}
\end{subequations}

Setting expressions in (\ref{NRQCD}) equal to those in
(\ref{FullQCDRe}) respectively, and expanding $\textrm{Im}f_n$ in
power of $\alpha_s$, we obtain the IR finite short distance
coefficients up to $\mathcal{O}(\alpha_{s}^{3})$:
\begin{subequations}
\begin{align}
2\textrm{Im}f(^{1}S_{0}^{[1]})=\frac{8\pi\alpha_s^2}{3}+\frac{\alpha_s^3}{27}(4(477-16N_f)
+12(33-2N_f)\ln(\frac{\mu^2}{4m_Q^2})-93\pi^2),
\end{align}
\begin{align}
2\textrm{Im}f(^{1}S_{0}^{[8]})=\frac{5\pi\alpha_s^2}{6}+\frac{5\alpha_s^3}{432}(16(153-4N_f)
+12(33-2N_f)\ln(\frac{\mu^2}{4m_Q^2})-129\pi^2),
\end{align}
\begin{align}
2\textrm{Im}f(^{1}P_{1}^{[8]})=\frac{\pi\alpha_s^2}{2}+\frac{\alpha_s^3}{108}\{-3(6N_f-61)\ln(\frac{\mu^2}{4m_Q^2})+2[-(24N_f+63\ln(2)+725)
\nonumber\\+119\pi^2+114\ln(\frac{\mu}{\mu_\Lambda})]\},
\end{align}
\begin{align}
2\textrm{Im}f(^{1}D_{2}^{[1]})=\frac{16\pi\alpha_s^2}{45}+\frac{\alpha_s^3}{1620}[78720-512N_f-7263\pi^2-4032\ln(2)
+2880\ln(\frac{\mu}{\mu_\Lambda})
\nonumber\\-192(N_f-9)\ln(\frac{\mu^2}{4m_Q^2})],
\label{Im-f:final}\end{align}
\end{subequations}
where the short distance coefficients of P-wave and D-wave are
$\mu_\Lambda$-dependent. Their $\mu_\Lambda$-dependence will be
canceled by that of the corresponding renormalized operators, which
could be obtained by finding the derivative of both sides of
(\ref{P8},\ref{S1},\ref{S8}) of $\mu_\Lambda$. For Born quantities,
$\frac{d\langle\mathcal{O}(^1L_{J}^{[1,8]})\rangle_{LO}}{d\mu_{\Lambda}}=0$.
Then we obtain the renormalization group equations at leading order
in $v$ and $\alpha_s$:
\begin{eqnarray}
\frac{d\langle\mathcal{O}^{R}(^1P_{1}^{[8]})\rangle_{NLO}}{
d\ln\mu_{\Lambda}}&=&\frac{8\alpha_sC_F}{3\pi
m_Q^2}\langle\mathcal{O}(^1D_{2}^{[1]})\rangle_{LO},\nonumber\\
\frac{d\langle\mathcal{O}^{R}(^1S_{0}^{[1]})\rangle_{NLO}}{
d\ln\mu_{\Lambda}}&=&\frac{1}{2N_c}\frac{8\alpha_s}{3\pi
m_Q^2}\langle\mathcal{O}(^1P_{1}^{[8]})\rangle_{LO},\nonumber\\
\frac{d\langle\mathcal{O}^{R}(^1S_{0}^{[8]})\rangle_{NLO}}{
d\ln\mu_{\Lambda}}&=&\frac{8\alpha_s B_F}{3\pi
m_Q^2}\langle\mathcal{O}(^1P_{1}^{[8]})\rangle_{LO}.
\end{eqnarray}
The solutions of the matrix elements $\langle
1{}^1D_{2}|\mathcal{O}(\mu_{\Lambda})|1{}^1D_{2}\rangle$ in heavy
quarkonium D-wave state ${}^1D_{2}$ are:
\begin{eqnarray}\label{evaluation}
\langle
1{}^1D_{2}|\mathcal{O}^{R}(^1P_{1}^{[8]})(\mu_{\Lambda})|1{}^1D_{2}\rangle
&=&\frac{8C_F}{3m_Q^2b_0}\ln\frac{\alpha_s(\mu_{\Lambda_0})}{\alpha_s(\mu_\Lambda)}\langle
1{}^1D_{2}|\mathcal{O}(^1D_{2}^{[1]})|1{}^1D_{2}\rangle,\nonumber\\
\langle
1{}^1D_{2}|\mathcal{O}^{R}(^1S_{0}^{[1]})(\mu_{\Lambda})|1{}^1D_{2}\rangle
&=&\frac{C_F}{4N_c}(\frac{8}{3m_Q^2b_0}\ln\frac{\alpha_s(\mu_{\Lambda_0})}{\alpha_s(\mu_\Lambda)})^2\langle
1{}^1D_{2}|\mathcal{O}(^1D_{2}^{[1]})|1{}^1D_{2}\rangle,\nonumber\\
\langle
1{}^1D_{2}|\mathcal{O}^{R}(^1S_{0}^{[8]})(\mu_{\Lambda})|1{}^1D_{2}\rangle
&=&\frac{C_FB_F}{2}(\frac{8}{3m_Q^2b_0}\ln\frac{\alpha_s(\mu_{\Lambda_0})}{\alpha_s(\mu_\Lambda)})^2\langle
1{}^1D_{2}|\mathcal{O}(^1D_{2}^{[1]})|1{}^1D_{2}\rangle,\nonumber\\
\end{eqnarray}
where the initial matrix elements like $\langle
1{}^1D_{2}|\mathcal{O}^{R}(^1P_{1}^{[8]})(\mu_{\Lambda_0})|1{}^1D_{2}\rangle$
at $\mu_{\Lambda_0}=m_Q\,v$ are eliminated at LO in
$v$~\cite{BBL,He}.

\section{numerical results and phenomenological discussions}
The long distance matrix element of D-wave four-fermion
color-singlet operator in the hadron state is related with the
second derivative of radial wave function at the origin through the
following relation:
\begin{equation}
\langle\label{H-D}
n{}^1D_2|\mathcal{O}(n{}^1D_2)|n{}^1D_2\rangle=\frac{15|R^{\prime\prime}_{nD}(0)|^2}{8\pi}=m_Q^6H_{Dn}
~ .
\end{equation}
Combining the leading order coefficient
$2\textrm{Im}f(^{1}D_{2}^{[1]})_{LO}=16\pi\alpha_s^2/45$ given in
(\ref{Im-f:final}) and the color-singlet matrix element given in
(\ref{H-D}), one can reproduce the decay width of ${}^1D_2$ state in
the CSM at leading order in $\alpha_s$~\cite{Novikov78}:
\begin{equation}
\Gamma_{CSM}(n{}^1D_2\to
gg)=\frac{2\alpha_s^2}{3}\frac{|R^{\prime\prime}_{nD}(0)|^2}{
m_Q^6}.
\end{equation}
However, there are contributions from the color-octet Fock states in
(\ref{FockState}) in NRQCD even at the order of $\alpha_s^2$. The
matrix elements of the P-wave octet operator and S-wave singlet as
well as octet operators in the ${}^1D_2$ bound state could be
estimated through the solutions of the operator evaluation equations
in (\ref{evaluation}).

\begin{figure}
\begin{center}
\includegraphics[bb=10 20 300 200]{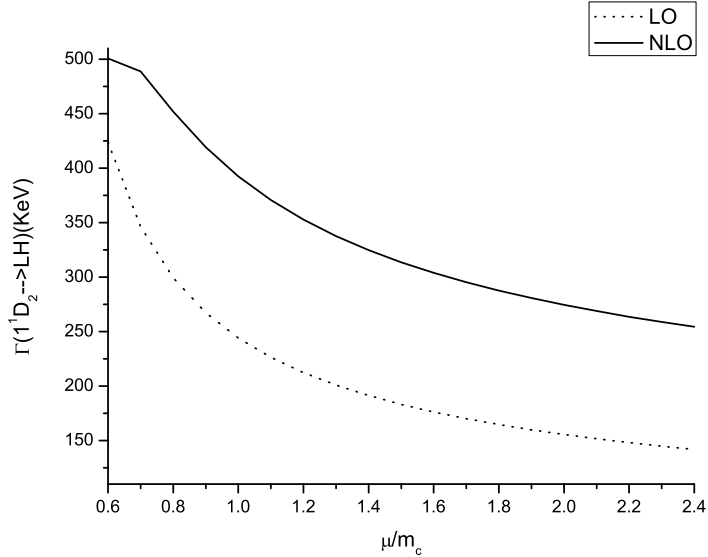}
\caption{Renormalization scale dependence of the decay width of
charmonium state $1^{1}D_{2}$ to LH }
\end{center}
\end{figure}

\begin{figure}
\begin{center}
\includegraphics[bb=10 20 300 200]{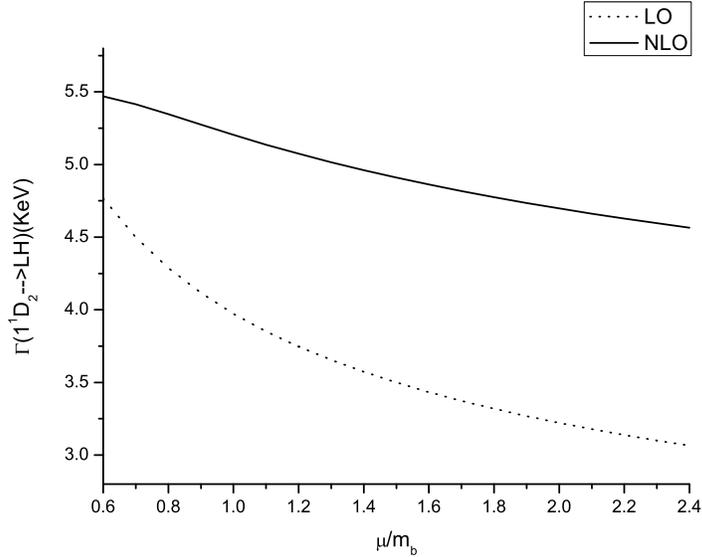}
\caption{Renormalization scale dependence of the decay width of
bottomonium state $1^{1}D_{2}$ to LH }
\end{center}
\end{figure}

The region of validity of the evolution equation is chosen as
follows: the lower limit $\mu_{\Lambda_0}=m_Q\upsilon$ and the upper
limit $\mu_\Lambda$ of order $m_Q$. For convenience, we take the
factorization scale $\mu_\Lambda$ to be the same as the
renormalization scale $\mu$ of order $m_Q$. We choose the pole mass
$m_c=1.5\textrm{GeV}$, $\upsilon^2=0.3$,
$\mu_{\Lambda_0}=m_c\upsilon$, $\mu_\Lambda=2m_c$,
$\alpha_s(2m_c)=0.249$, $N_f=3$, $\Lambda_{QCD}=390\textrm{MeV}$,
$H_{D1}=\frac{15|R^{\prime\prime}_{1D}(0)|^2}{8\pi
m_c^6}=0.786\times10^{-3}\textrm{GeV}$~\cite{Wavefunction} for
charmonium, and
$m_b=4.6\textrm{GeV},\upsilon^2=0.1,\mu_{\Lambda_0}=m_b\upsilon,\mu_\Lambda=2m_b,
\alpha_s(2m_b)=0.180,N_f=4,\Lambda_{QCD}=340\textrm{MeV}$,
$H_{D1}=\frac{15|R^{\prime\prime}_{1D}(0)|^2}{8\pi
m_b^6}=0.401\times10^{-4}\textrm{GeV}$ for $1D$ states and
$H_{D2}=\frac{15|R^{\prime\prime}_{2D}(0)|^2}{8\pi
m_b^6}=0.750\times10^{-4}\textrm{GeV}$ for $2D$
states~\cite{Wavefunction} for bottomonium. The $\mu$ dependence
curves of the decay widths are shown in Fig. 9 and Fig. 10. When
$\mu=2m_{c}$ for $c\bar{c}$ systems and $2m_{b}$ for $b\bar{b}$
systems, we get the predictions at $\mathcal{O}(\alpha_s^3)$:
\begin{eqnarray}
\Gamma_C(1^1D_2\rightarrow LH)&=&274\textrm{KeV},\nonumber\\
\Gamma_B(1^1D_2\rightarrow LH)&=&4.70\textrm{KeV},\nonumber\\
\Gamma_B(2^1D_2\rightarrow LH)&=&8.78\textrm{KeV}.
\end{eqnarray}
The LO decay widths for charmonium and bottomonium $1{}^1D_2$ states
are 155 KeV and 3.22 KeV. Therefore the NLO QCD corrections
contribute enhancement of factor 1.8 and 1.5, respectively.

\begin{center}
\begin{table}
\caption{Subprocess decay rates of $1{}^1D_2$ charmonium, where
$m_c=1.5\textrm{GeV}$, $\upsilon^2=0.3$,
$\mu_{\Lambda_0}=m_c\upsilon$, $\mu_\Lambda=2m_c$ and
$\alpha_s(2m_c)=0.249$.}
\begin{tabular}{|c|c|c|c|}
     \hline
     Subprocess & LO(KeV) & NLO(KeV)\\
     \hline
     $({}^1D_2)_1\rightarrow LH$  & 54.7 & 75.1 \\
     \hline
     $({}^1P_1)_8\rightarrow LH$  & 66.6 & 132 \\
     \hline
     $({}^1S_0)_8\rightarrow LH$  & 15.0 & 31.3 \\
     \hline
     $({}^1S_0)_1\rightarrow LH$  & 19.2 & 36.1 \\
     \hline
\end{tabular}
\end{table}
\end{center}

For the $1{}^1D_2$ charmonium state $\eta_{c2}$, the numerical
values for all subprocesses are also listed in Table~V. One can see
from this table that the contributions from the Fock states other
than $|{^1D_2^{[1]}}\rangle$ are dominant in the decay width, and
the total result is about 1-3 times larger than that in CSM even at
leading order in $\alpha_s$.

For the phenomenological analysis of $\eta_{c2}$, we vary the
renormalization/factorization scale from $2m_c$ to $m_c$ and get
$\Gamma(\eta_{c2}\to LH)=274\mbox{-}392$ KeV. The electric
transition rate $\Gamma(\eta_{c2}\to\gamma h_{c})=339\mbox{-}375$
KeV~\cite{Barnes} and the dipion transition rate
$\Gamma(\eta_{c2}\to \eta_{c}\pi\pi)\approx45$ KeV~\cite{Eichten}
have been estimated elsewhere in the literature.

As emphasized before, the $\eta_{c2}$ should be a narrow state,
since its mass and quantum numbers forbid it to decay into charmed
meson pairs $D\bar{D}$ and  $D^*\bar{D}$. Therefore, the main decays
modes of $\eta_{c2}$ are expected to be the electric as well as
hadronic transitions to lower-lying charmonium states and the
inclusive light hadronic decay. With all these decay widths given
above, we get the total width of $\eta_{c2}$ to be about
660-810~KeV, and the branching ratio of the electric transition to
be
\begin{equation}
\mathcal{B}(\eta_{c2}\to\gamma h_{c})=(44\,\mbox{-}\,54)\%,
\end{equation}
which provides important information on probing this missing state.
In practice, one can search for $\eta_{c2}$ through the cascade
decay $\eta_{c2}\to \gamma h_c\to \gamma\gamma \eta_c\to
\gamma\gamma K\bar K\pi$ with branching ratios
$\mathcal{B}(h_{c}\to\gamma \eta_{c})\approx 0.4$~\cite{Eichten,
chao1993} and $\mathcal{B}(\eta_{c}\to K\bar
K\pi)\approx7\%$~\cite{PDG}. Similar decay chains can also be used
to search for the $\eta_{b2}^{(\prime)}$.

The production rates of $\eta_{c2}$ are expected to be generally low
in many processes, because the rates are suppressed by the small
values of the second derivative of the wave function at the origin,
and also by its spin-singlet nature, which forbids $\eta_{c2}$ to
couple to a photon, or to be detected from the E1 transitions of
higher spin-triplet charmonia. Nevertheless, efforts should be made
to find this very unique missing charmonium state. Hopefully, the
study for the inclusive light hadronic decay of $\eta_{c2}$ in NRQCD
will provide useful information on searching for this state in
high-energy $p\bar p$ collision~\cite{Cho}, in $B$ decays~\cite{Ko},
in higher charmonium transitions, in $e^+e^-$ process in BESIII at
BEPC~\cite{BES3}, and particularly in the low-energy $p\bar p$
reaction in PANDA at FAIR~\cite{PANDA}.




\section{Summary}
In this paper, we calculate the inclusive light hadronic decay width
of the ${}^1D_2$ heavy quarkonium state up to order of $\alpha_s^3$
and at the leading order in $v$ within the framework of NRQCD. We
find that the inclusive decay widths into light hadrons via gluons
and light quarks at order of $\alpha_s^3$ in QCD suffer from both IR
divergences and Coulomb singularities, but they can be absorbed into
the renormalization of the matrix elements of the four-fermion
operators in NRQCD precisely. Therefore, after matching the full QCD
onto NRQCD, the IR divergent part is removed, and IR finite
short-distant coefficients are obtained, and the dependence on the
factorization scale of the coefficient is canceled by that of the
corresponding matrix element with the renormalization group
analysis.

At leading order in $\alpha_s$, the result in the CSM can be
reproduced but there are many other contributions, such as that from
color-octet P-wave operators, which will enhance the width in CSM by
several times in magnitude even at the leading order in $\alpha_s$.
Furthermore, the NLO results give extra enhancement factors of 1.8
for $\eta_{c2}$ and 1.5 for $\eta_{b2}$ relative to the LO ones,
respectively. By choosing the factorization scale as $2m_Q$, the
light hadronic decay widths are found to be about 274, 4.7, and 8.8
KeV for the $\eta_{c2},~ \eta_{b2}$, and $\eta_{b2}'$ respectively.
Based on these estimates, and using the E1 transition width and
dipion transition width for the $\eta_{c2}$ estimated elsewhere in
the literature, we get the total width of $\eta_{c2}$ to be about
660-810~KeV, and the branching ratio of the electric transition
$\eta_{c2}\to\gamma\,h_c$ to be about $(44\,\mbox{-}\,54)\%$, which
will be useful in searching for this missing charmonium state
through, e.g., the process $\eta_{c2}\to\gamma\,h_c$ followed by
$h_c\to\gamma\eta_c$.


\section{Acknowledgement}
Y.F. would like to thank Dr. Ce Meng and Dr. Yu-Jie Zhang for useful
discussion and reading the manuscript. Y.F. would also like to thank
Mr. Rolf Mertig and Prof. Fabio Maltoni for their useful suggestions
by e-mail. Z.G.H. thanks Institute of High Energy Physics of Chinese
Academy of Sciences and Theoretical Physics Center for Science
Facilities for their hospitality. This work was supported in part by
the National Natural Science Foundation of China (No 10675003, No
10721063).

\end{document}